\documentclass{aa}  
\usepackage[utf8]{inputenc}
\usepackage{amssymb,amsmath}
\usepackage{graphicx}
\usepackage[dvipsnames]{xcolor}
\usepackage[breaklinks=true]{hyperref}
\hypersetup{colorlinks=true,allcolors=teal}
\usepackage[flushleft]{threeparttable}
\usepackage{footmisc}
\usepackage{subfigure}
\usepackage{multirow}
\usepackage{placeins}
\newcommand{\be}{\begin{equation}}
\newcommand{\ee}{\end{equation}}
\def\bear#1\ear{\begin{align}#1\end{align}}

\newcommand{\tab}[1]{Table~\ref{#1}}

\usepackage[T1]{fontenc}
\usepackage{amsmath}	
\usepackage{orcidlink}
\DeclareRobustCommand{\VAN}[3]{#2}
\let\VANthebibliography\thebibliography
\def\thebibliography{\DeclareRobustCommand{\VAN}[3]{##3}\VANthebibliography}
\usepackage{graphicx}
\usepackage{txfonts}

\begin{document}

   \title{Studying dark gaps in Ly-$\alpha$ forest transmission with large reionization simulations}
    \authorrunning{Maity et al.}

   \author{Barun Maity\inst{1}\fnmsep\thanks{maity@mpia.de}\orcidlink{0000-0002-4682-6970}
          ,
          Frederick B. Davies
          \inst{1}\orcidlink{0000-0003-0821-3644},
          Benedetta Spina \inst{2}\orcidlink{0000-0003-1634-1283} \and
          Sarah E. I. Bosman \inst{2}\orcidlink{0000-0001-8582-7012}}

   \institute{Max-Planck-Institut f\"ur Astronomie, K\"onigstuhl 17, D-69117 Heidelberg, Germany
   \and Institute for Theoretical Physics, Heidelberg University, Philosophenweg 12, D–69120, Heidelberg, Germany}
   
   \date{Received XXX; accepted XXX}
   
  \abstract{The physical conditions of the intergalactic medium (IGM) during the final stages of cosmic reionization ($z\sim5.0-6.0$) are not yet fully understood. Recent reports of unexpectedly large-scale ($\ge 150 h^{-1}\mathrm{cMpc}$) correlation in Ly-$\alpha$ transmission flux using extended XQR-30 quasar spectra pose interesting challenges on the reionization end stages. In this work, we investigate the Ly-$\alpha$ forest dark-gap distribution (defined as regions with transmitted flux below 0.05) as another sensitive tracer of the IGM, using an efficient, large-volume ($\sim 1 ~\mathrm{Gpc}$) simulation framework. By constructing a suite of physically motivated model variants (i.e, varying the reionization redshift, IGM temperature, and ionizing-photon mean free path), we generate synthetic sightlines and compare their predicted cumulative distribution of dark gaps with that of observed spectra (at redshift intervals of $\Delta z=0.2$). We find that most of the models achieve qualitatively consistent agreement with the data. The scenario involving a slightly later reionization completion ($z\sim 5.4$) provides the closest match, while a short constant mean free path model is disfavored by the data at lower redshifts. These findings give qualitative support for the emerging scenario of reionization end extending to $z\le5.7$, although they can not rule out a slightly early reionization with enhanced post-ionization ultraviolet (UV) background fluctuations. A similar conclusion arises from the redshift distribution of long dark gap ($L\ge 30 ~h^{-1}\mathrm{cMpc}$) fraction. However, the model variants are still not able to reproduce the observed strong flux correlations at unusually large scales, which remains open for further investigations.
}

   \maketitle
   
\section{Introduction}
\label{sec:intro}
The Epoch of Reionization (EoR) forms a critical bridge between the early and present-day Universe, preserving the imprints of the first luminous sources. During this epoch, the collective radiation from those sources drove the transformation of the intergalactic medium (IGM) from an almost entirely neutral state following recombination to a highly ionized one. Although substantial theoretical and observational advances have been made, the detailed morphology, duration, and physical processes governing the EoR remain among the most significant open questions in contemporary cosmology \citep[see reviews for details,][]{2001PhR...349..125B,2009CSci...97..841C,2016ARA&A..54..313M, 2018PhR...780....1D,2022arXiv220802260G,2022GReGr..54..102C}.

A diverse set of observational probes has been developed to constrain the neutral fraction of the IGM during the reionization era. These include measurements based on the damping wing signatures observed in quasar \citep[e.g.,][]{Davies18,2019MNRAS.484.5094G,2022MNRAS.512.5390G, Greig2024, Wang2020} and galaxy \citep[e.g.,][]{Umeda2024,2025arXiv250404683U} spectra, Lyman (Ly)-$\alpha$ emitters luminosity functions \citep[e.g.,][]{Morales2021,2025ApJS..278...33K} and their clustering properties \citep[e.g.,][]{Sobacchi2015,2025ApJS..277...37U}, as well as Ly-$\alpha$ equivalent widths \citep[e.g.,][]{2018ApJ...856....2M,2025MNRAS.536.2355J}, collectively probing redshifts up to $z\lesssim12.0$. These studies generally point to a midpoint of reionization $z_{\mathrm{re}}\sim7.0-8.0$ \citep{2019ApJ...878...12H,2018Natur.553..473B}, although the derived constraints are often limited by uncertainties in continuum modeling of the underlying spectra. Complementarily, at relatively lower redshifts ($z\lesssim6.0$),  more robust constraints have emerged from dark-pixel fraction analyses \citep{2023ApJ...942...59J,davies26} and from damping-wing signatures identified in stacked Ly-$\alpha$ forest spectra \citep{2024A&A...688L..26S,Zhu2024_damping}. Moreover, the pronounced large-scale fluctuations observed in the  Ly-$\alpha$ forest effective opacity distribution ($\tau_{\mathrm{eff}}$) suggest that the reionization may extend towards lower redshifts, even below $z<5.5$ \citep{2019MNRAS.485L..24K, Bosman22}.

A range of theoretical models has been proposed to account for the excess fluctuations observed in the Ly-$\alpha$ forest opacity at lower redshifts. Under the assumption that the IGM is already highly ionized in this regime, these fluctuations may arise from residual inhomogeneities in the temperature field \citep{2015ApJ...813L..38D} and/or from spatial variations in the ultraviolet background (UVB) driven by a relatively short ionizing-photon mean free path \citep{Davies16, Maity_davies_2025}. Alternatively, the most widely discussed explanation invokes the presence of residual neutral islands, implying that reionization concluded later than traditionally assumed \citep{2020MNRAS.497..906K,2020MNRAS.494.3080N,2021MNRAS.501.5782C,Qin2025}. This scenario is also supported by the rapid evolution of photon mean free path at $6.0\geq z\geq5.0$ \citep{2021MNRAS.508.1853B}, as well as large-scale underdensities associated with long dark troughs \citep[e.g., $\sim 110 h^{-1}\mathrm{cMpc}$ of ULAS J0148+0600;][]{2015MNRAS.447.3402B} traced by the spectra of Ly-$\alpha$ emitters \citep{2023ApJ...955..138C} and Ly-break galaxies \citep{2020ApJ...888....6K}. Nevertheless, these observations do not rule out an early-reionization scenario characterized by substantial UVB fluctuations \citep{Zhu_2021}.

Therefore, key questions concerning the influence and spatial extent of neutral islands at these lower redshifts remain unresolved. Improved measurements of Ly-$\alpha$ forest opacity fluctuations and their redshift evolution may offer a promising avenue for breaking the degeneracy among competing models. Additional complementary constraints can be obtained from the statistics of dark gaps (defined as contiguous regions where the transmitted flux falls below a specified threshold) in the forest, which may arise either from persistent neutral patches or from regions permeated by a relatively weak UV background \citep[e.g.,][]{2006AJ....132..117F,2008MNRAS.386..359G,2017ApJ...841...26G,2020MNRAS.494.3080N}. Recent efforts involving dark gaps statistics in Ly-$\alpha$ forest also indicate strong degeneracy between these models \citep{Zhu_2021}. However, earlier studies were limited by the modest sizes of the simulation volumes employed ($\le 160~ h^{-1}\mathrm{cMpc}$), constrained by computational cost and model complexity \citep[e.g.,][]{2017ApJ...841...26G,Zhu_2021}. More recently, an unexpectedly large-scale correlation in the Ly-$\alpha$ forest ($\ge 150~h^{-1}\mathrm{cMpc}$) has been reported \citep{Spina25} using the extended (E)-XQR-30 dataset \citep{DOdorico23}. This finding motivates the use of substantially larger simulation volumes, such as the BIBORTON box ($\sim 1024 ~h^{-1}\mathrm{cMpc}$), capable of capturing fluctuations on the relevant scales, in combination with the efficient  Ly-$\alpha$ forest modeling framework developed by \citet{Maity_davies_2025}. Nevertheless, the fiducial large-box model fails to reproduce the observed large-scale correlations within the range of currently understood physical mechanisms \citep{Spina25}.  

To address these issues, we extend our investigation by exploring a suite of model variants within large-volume simulations. In particular, we examine scenarios that differ in the timing of reionization completion, the amplitude of UVB fluctuations, and the thermal state of the IGM. Our goal is to evaluate the performance of these models against the other known observables derived from Ly-$\alpha$ forest, such as statistics using dark pixel distributions at the redshift range $5.0\le z \le 6.1$. We utilized 42 high S/N quasar spectra from the  E-XQR-30 dataset, after carefully avoiding the contamination due to damped Ly-$\alpha$ (DLA) and O-VI lines, and compared them with our model predictions. This study thus provides an important robustness test of the different model realizations, thereby reinforcing the interpretations suggested by current observational findings.

The paper is organized as follows. In Section \ref{sec:sim_brief}, we briefly discuss the simulation methodology, introducing different variants of the model parameters. We lay out the data reduction procedure to get the dark pixel distribution in Section \ref{sec:obs_dat}. Following this, we discuss the statistics that have been studied in Section \ref{sec:stat}. We summarize our findings and interpretations subsequently in the Section \ref{sec:results}. Finally, we conclude the work in Section \ref{sec:summary}. Throughout this paper, we use $h^{-1}\mathrm{cMpc}$ as distance unit unless otherwise stated and adopt \citet{2020A&A...641A...6P} values of cosmological parameters ($\Omega_m=0.308$, $h=0.678$, $\Omega_{\mathrm{\Lambda}}=0.691$).

\begin{figure*}
    \centering
    \includegraphics[width=\textwidth]{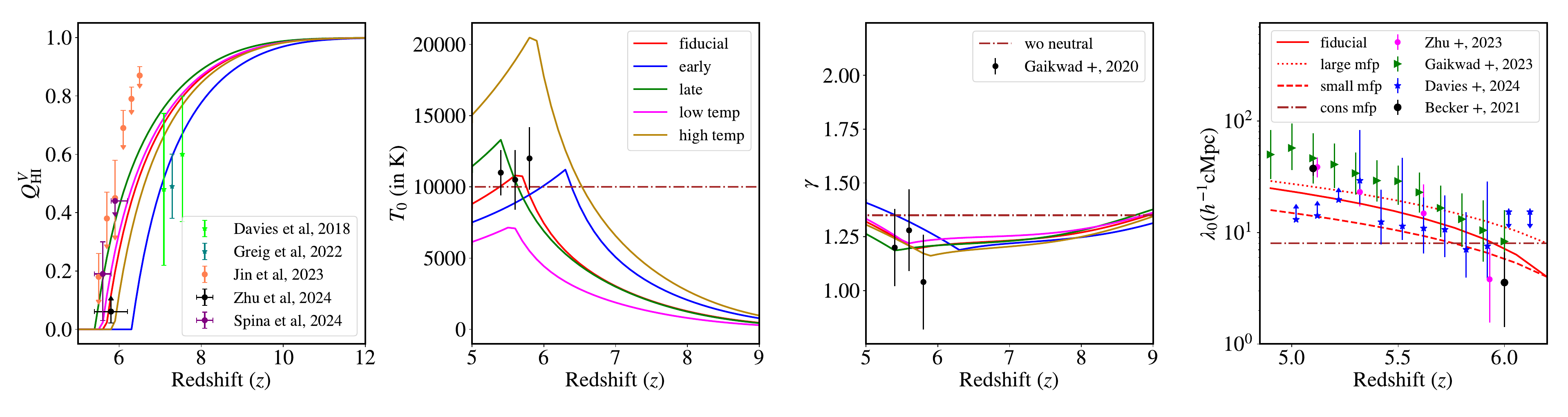}
    \caption{Different observables for the various model scenarios assumed in this study. From left to right, the panels show redshift evolution of global neutral fraction ($Q_{\mathrm{HI}}^V$), mean IGM temperature ($T_0$), index of temperature-density relation ($\gamma$) and the effective photon mean free path ($\lambda_0$). We also show various constraints on these quantities, as suggested by recent studies, i.e, constraints on neutral fraction \citep[][]{Davies18, 2022MNRAS.512.5390G,2023ApJ...942...59J, 2024A&A...688L..26S, Zhu2024_damping}, IGM temperature estimates \citep[][]{2020MNRAS.494.5091G}, mean free path estimates \citep[][]{2021MNRAS.508.1853B,Gaikwad23,2023ApJ...955..115Z, Davies24}. Note that, ``cons mfp'' model assumes $Q^   V_{\mathrm{HI}}=0$ throughout the redshift ranges.}
    \label{fig:model_var}
\end{figure*}

\begin{figure*}
    \centering
    \includegraphics[width=\textwidth]{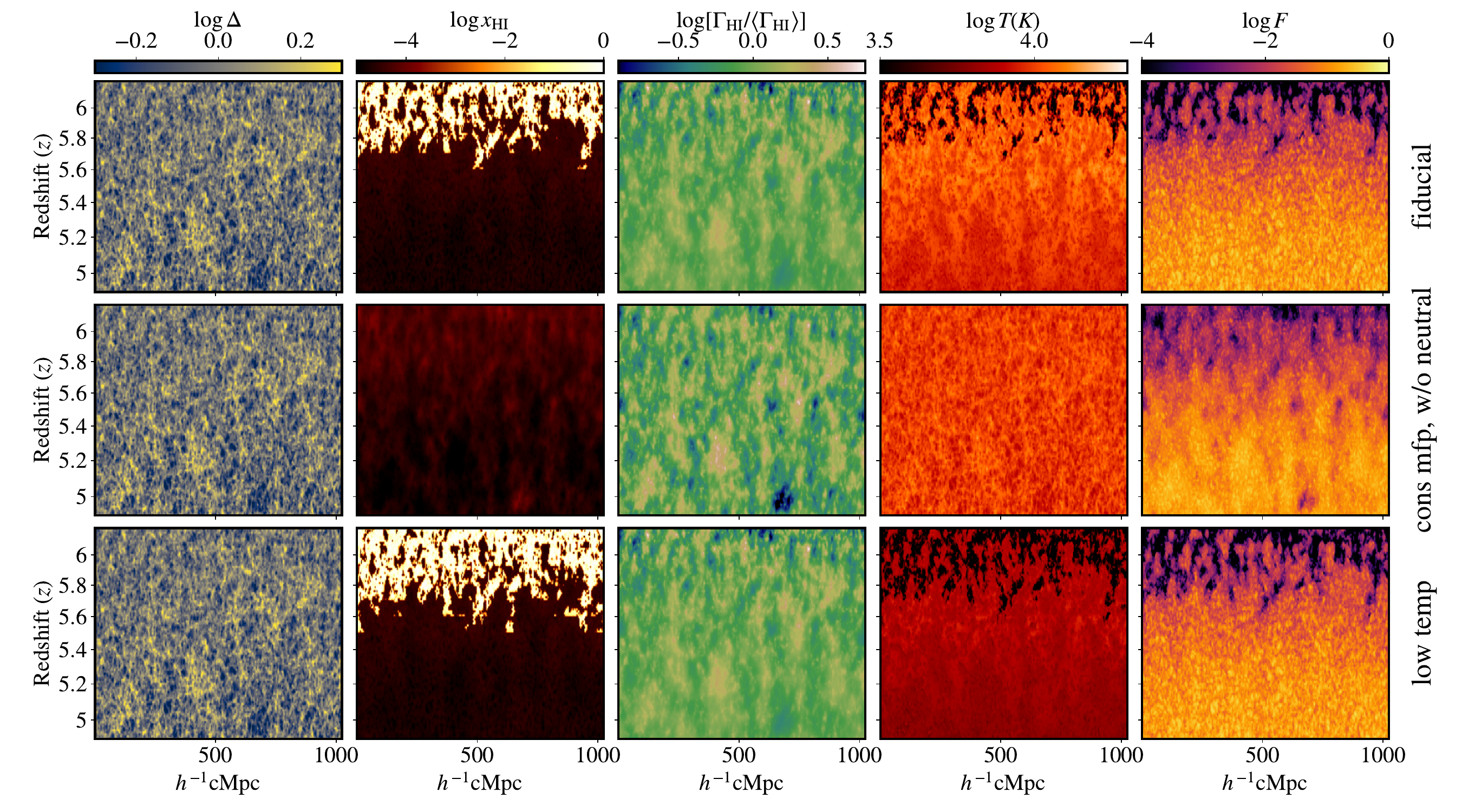}
    \caption{Lightcone snapshots for three different cases in (\textit{top}: fiducial, \textit{middle}: cons mfp, w/o neutral, and \textit{bottom}: low temp) in three rows. The columns correspond to density ($\Delta$), neutral fractions ($x_{\mathrm{HI}}$), UVB fluctuations ($\Gamma_{\mathrm{HI}}/\langle \Gamma_{\mathrm{HI}}\rangle$), temperature ($T$), and flux ($F$). The colorbars have been shown in logarithmic scales. The rest of the scenarios has been shown in Appendix \ref{app:appendix2}.}
    \label{fig:mdel_snap}
\end{figure*}

\section{Simulation in brief}\label{sec:sim_brief}
\subsection{Methodology}\label{sec:method}
\begin{figure*}
    \centering
    \includegraphics[width=\textwidth]{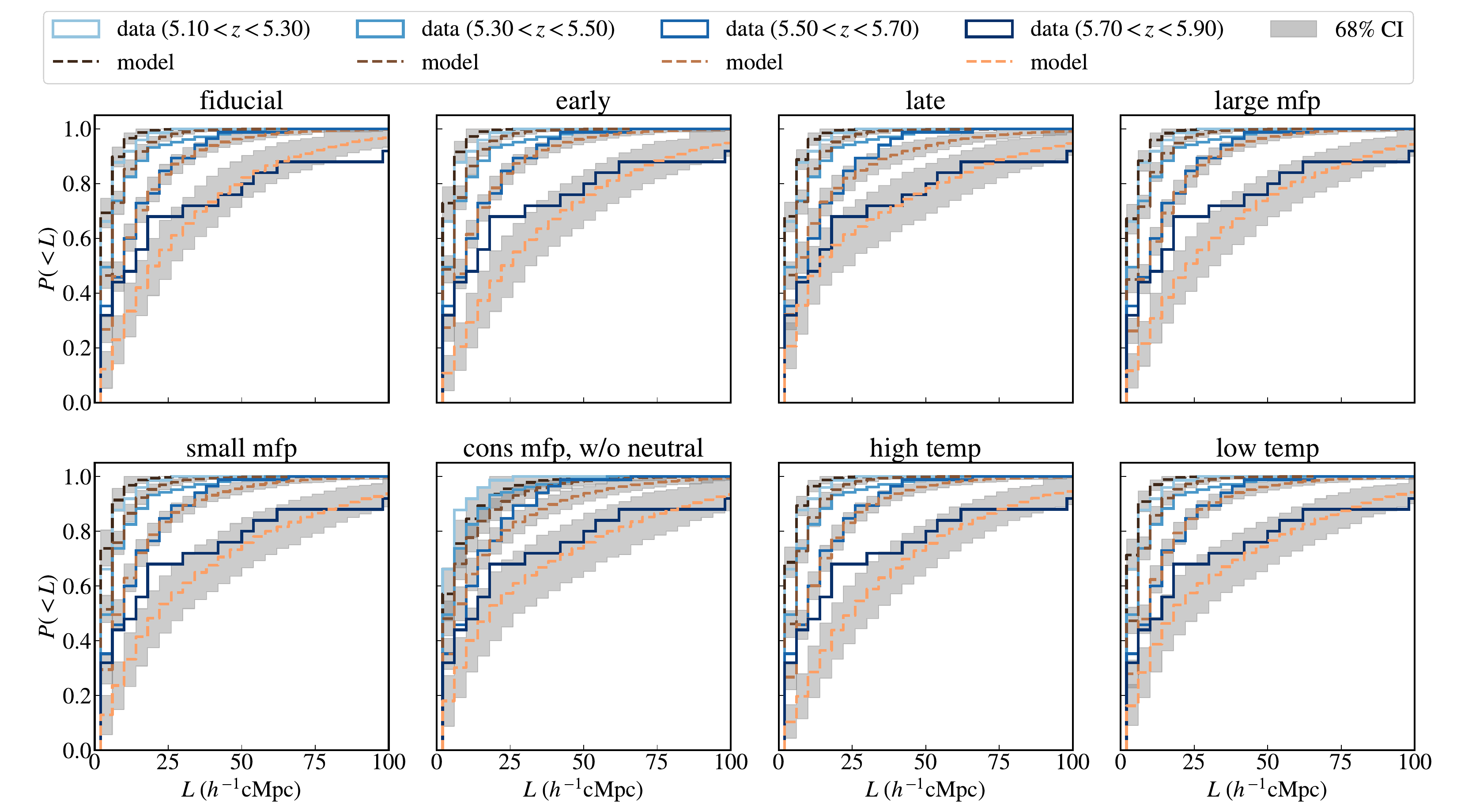}
    \caption{The cumulative probability distribution functions (CDFs) of dark gap lengths at different redshift ranges (within an interval of $\Delta z=0.2$), corresponding to the set of different model variants, discussed in section \ref{sec:model_suite}. The \textit{solid} lines denote the derived distribution from the observed data, while the \textit{dashed} lines are the corresponding predictions using model skewers. The shaded region signifies 68\% uncertainties on the model distributions.}
    \label{fig:dark_cdf}
\end{figure*}
The simulation methodology utilized in this study has been exploited and described in a recent study by \citet{Spina25}. The discussion in this section closely follows the simulation description of that earlier work. To better understand the physical mechanisms at play during the late EoR, we aim to compare our observational results with physically motivated simulations of the intergalactic medium IGM. Ideally, this would require a simulation volume with both high dynamic range, preserving large-scale correlations, while also resolving small-scale physics. However, achieving this level of detail is computationally prohibitive with current resources.

As an alternative, we adopt a complementary strategy, applying a semi-numerical technique developed by \citet{Maity_davies_2025}. This method efficiently produces large-volume lightcones of Ly$\alpha$ transmission across the redshift range $z = 4.9$–$6.2$ (spanning a comoving length of approximately $420~h^{-1}\mathrm{cMpc}$), which is well-suited to our study.

In our semi-numerical framework, the Ly-$\alpha$ optical depth is modeled as a function of underlying cosmological density fluctuations, UVB variations, temperature inhomogeneities, and ionization fluctuations (where reionization remains incomplete). These relationships are calibrated against a high-resolution, fully hydrodynamic simulation—specifically, the \texttt{Nyx} code \citep{Almgren13}. The semi-numerical UVB and reionization models additionally require the collapsed halo mass fraction ($f_{\mathrm{coll}}$) to estimate the available ionizing photon budget. For this purpose, we generate density fields at fixed redshift using the Zel’dovich approximation \citep{zeldovich70} in a large computational volume ($L = 1024~h^{-1}\mathrm{Mpc}$). The collapsed mass fraction field is calculated in Lagrangian space via the Excursion Set Formalism \citep[ESF-L;][]{Trac22}. This combination reproduces large-scale structure consistent with full N-body simulations while allowing efficient exploration of EoR evolution. 

The ionization and temperature histories are then evolved using the photon-conserving reionization model SCRIPT, which includes recombination effects \citep{Maity22}. In this model, the temperature increment due to photoionization heating is associated with a free parameter, $T_{\mathrm{re}}$, and the avilable photon budget is determined by the ionizing efficiency, $\zeta$ (assuming a power-law variation with redshift). For computational practicality, we do not explicitly model radiative feedback; instead, its influence is approximated by enforcing a fixed minimum halo mass ($M_{\mathrm{min}} = 10^9 M_{\odot}$), below which structure formation is suppressed.

Our simulations employ a spatial resolution of $\Delta x = 4~h^{-1}\mathrm{Mpc}$, balancing computational efficiency and accuracy. To track lightcone evolution, we produce simulation snapshots at intervals of $\Delta z = 0.1$, interpolating between them as needed. Given the density and source fields, we apply the EX-CITE model \citep{Gaikwad23} to generate UVB fluctuations, incorporating local source contributions as described in \citet{Davies16,Davies24}. This process is controlled by two key parameters: the effective mean free path of ionizing photons ($\lambda_\mathrm{mfp}$) and the average photoionization rate ($\langle\Gamma_{\mathrm{HI}}\rangle$). All these quantities are utilized to get the transmission flux lightcones via the calibrated Fluctuating Gunn Peterson Approximation (FGPA) as detailed in \citet{Maity_davies_2025}. We tune the mean photoionization rate for each coeval boxes such that the mean transmission flux match the estimates of \citet{Bosman22}.

\subsection{Model suite}\label{sec:model_suite}
In this section, we describe the suite of models employed to compute the dark-pixel statistics. The models are outlined below:
\begin{itemize}
    \item fiducial: This model corresponds to a realistic scenario where reionization ends at $z\sim5.6-5.7$ and the IGM temperatures are modelled in a self-consistent way as described in \citet{Maity_davies_2025}. This corresponds to a temperature increment parameter ($T_{\mathrm{re}}$) of $10^{4.2}~K$. The mean free path parameter ($\lambda_0$) evolves from $4~h^{-1}\mathrm{cMpc}$ at $z=6.2$ to $25~h^{-1}\mathrm{cMpc}$ at $z=4.9$. The fiducial model obeys the recent observational constraints on ionization and temperature evolution.
    \item early: This is a variant of the fiducial model in which reionization ends earlier, at $z\sim6.3$. The earlier completion is achieved by increasing the ionizing-efficiency parameter, $\zeta$, while retaining the self-consistent thermal and ionization evolution. This has the same mean free path evolution as the fiducial case.
    \item late: This corresponds to a complementary variant where reionization ends slightly late (at $z\sim5.4$). This is realized by reducing the ionizing-efficiency parameter relative to the fiducial case. The mean free path evolution is again kept the same as the fiducial scenario.
    \item large mfp: This model is designed to explore reduced UVB inhomogeneity, achieved by increasing the mean free path at every redshift by $4~h^{-1}\mathrm{cMpc}$ relative to the fiducial values (same reionization end as fiducial). 
    \item small mfp: This scenario adopts a reduced mean free path than the fiducial, thereby increasing UVB fluctuations. Specifically, $\lambda_0$ evolves from $15~ h^{-1}\mathrm{cMpc}$ at $z=5.0$ to $4~h^{-1}\mathrm{cMpc}$ at $z=6.2$. This scenario is also closer to the estimates of \citet{Maity_davies_2025}, based on Ly-$\alpha$ forest opacity fluctuations at the intermediate redshifts ($z\sim5.4-5.7$). Similar to previous, this also adopts the same ionization evolution as the fiducial model. 
    \item cons mfp, w/o neutral: In this model, no residual neutral islands ($Q_{\mathrm{HI}}\sim0$) are assumed to exist in the targeted redshift interval, and additionally, no temperature evolution is introduced. The IGM temperature follows a standard equation of state, $T=T_0\Delta^{\gamma-1}$, with $T_0=10^4~K$ and $\gamma=1.35$. We also use a constant $\lambda_0$ of $8~h^{-1}\mathrm{cMpc}$ throughout the redshift range instead of evolution. 
    \item high temp: This corresponds to a high-heating scenario in which the $T_{\mathrm{re}}$ parameter is set to $3\times10^4~K$, which is substantially higher than the fiducial case. This corresponds to a slightly earlier reionization end than the fiducial due to the reduced strength of recombination.
    \item low temp: This is a complementary version of the previous case, with cooler IGM (fixing $T_{\mathrm{re}}$ at $10^4~K$), which is considerably smaller than the fiducial scenario. This pushes the reionization end towards slightly lower redshift than the fiducial scenario due to stronger recombination.
\end{itemize}

In Figure \ref{fig:model_var}, we present the ionization histories, temperature evolution, and mean free path evolution for all model variants described above. As expected, the fiducial model satisfies the available observational constraints, while the alternative variants diverge in ways consistent with their construction. For instance, the early reionization model doesn't obey the nearly model-independent constraints on neutral fraction \citep{2023ApJ...942...59J, 2024A&A...688L..26S}. The models with lower (higher) temperature produce an IGM  temperature ($T_0$) well below (above) the estimates of \citet{2020MNRAS.494.5091G}. These deviations highlight the parameter sensitivity of the thermal and ionization histories and their probable influence on the predicted Ly-$\alpha$ forest observables. 

In Figure \ref{fig:mdel_snap}, we further illustrate the spatial distribution of the key physical quantities through snapshots (i.e., neutral fraction, UVB fluctuations, temperature, and transmission flux) corresponding to three representative model variants (i.e., fiducial, cons mfp, w/o neutral, and low temp, from top to bottom row). Notably, lowering the temperature has only a minimal impact on the transmitted-flux morphology (i.e., top and bottom rows), without strongly affecting large-scale opacity. This behaviour is also expected from the fact that the opacities for all the models have been scaled to match the mean transmitted flux, countering the effect of temperature modifications at large scales. By contrast, the absence of mean free path evolution or neutral island (i.e., middle row) can produce a substantially different flux field. Specifically, the flux field is much smoother without the effect of neutral island and can produce dark regions that persist even at a redshift of $z\sim5.0$ due to existing strong UVB fluctuations (as a consequence of relatively short mean free path). 

These comparisons emphasize that certain physical ingredients, such as the presence of residual neutral islands and the redshift evolution of the mean free path, play a dominant role in shaping large-scale Ly-$\alpha$ forest opacity fluctuations. Consequently, dark-pixel statistics may provide a sensitive diagnostic for distinguishing among these scenarios.
\begin{figure*}
    \centering
    \includegraphics[width=0.97\textwidth]{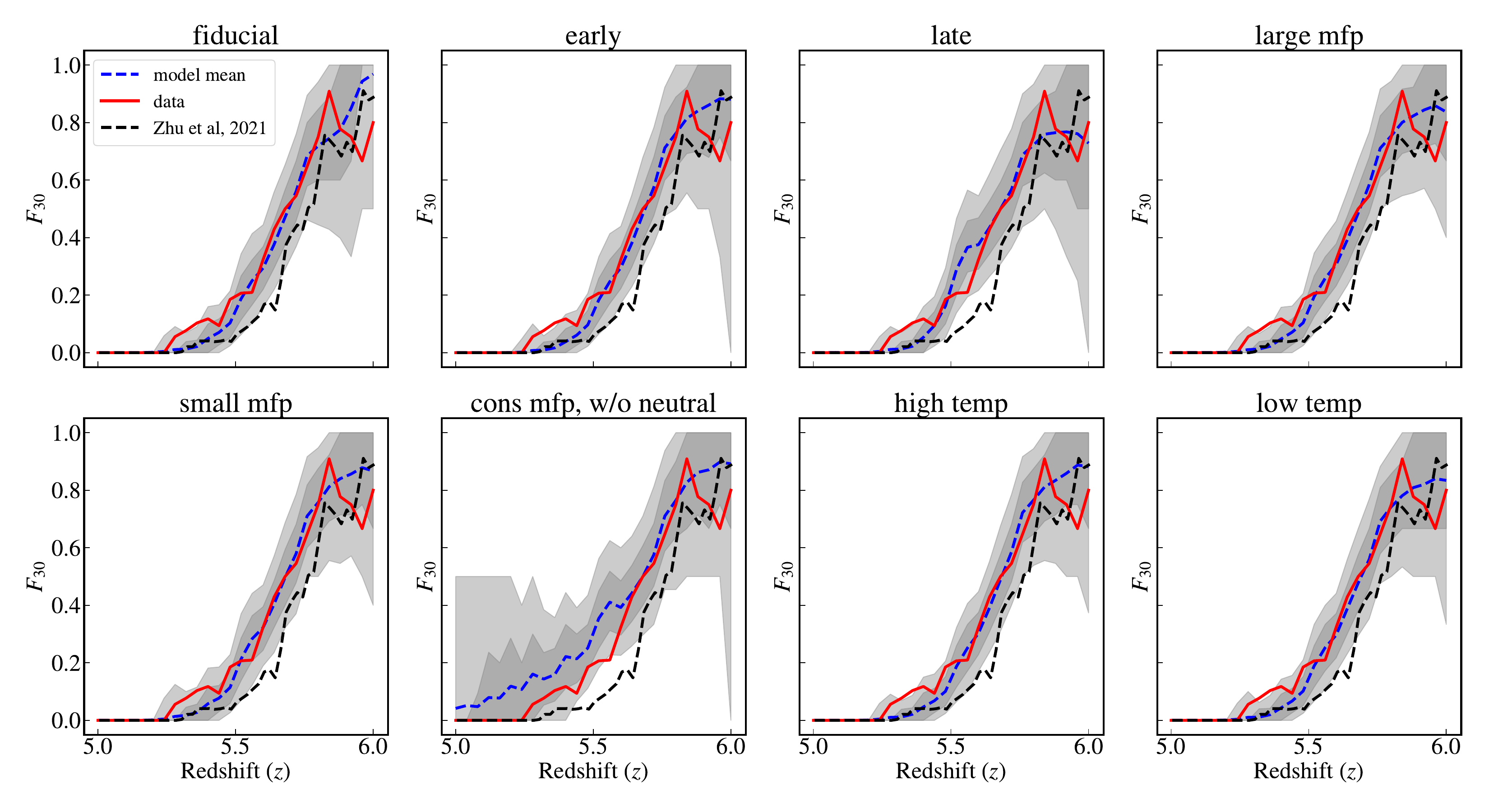}
    \caption{Fraction of skewers with dark gap length, $L\ge 30 ~h^{-1}\mathrm{cMpc}$ ($F_{30}$) as a function of redshift ($z$) for the different model variants, discussed in section \ref{sec:model_suite}. The \textit{red} lines are derived distributions from the observed spectra. The \textit{blue} dashed lines denote the mean distribution after averaging over skewer realizations from the model suites. The shaded regions show the corresponding 68\% and 95\% uncertainties. In \textit{black} dashed, we show similar estimates from an earlier study \citep{Zhu_2021}, with different skewer resolution and samples.}
    \label{fig:F30_z}
\end{figure*}

\section{Observational data}
\label{sec:obs_dat}
We use a sample of the observational dataset described in \citet{Spina25} implementing the reduction procedures, consisting of the 42 high-redshift quasar spectra at $z > 5.5$ with $\mathrm{S/N} \geq 10$ per $\leq 15~\mathrm{km~s^{-1}}$ pixel of the E-XQR-30 sample \citep{Bosman22,DOdorico23}. The spectra were obtained with VLT/X-Shooter \citep{Vernet11} and reduced following the procedures outlined in \citet{Bosman22}. Each sightline is continuum-normalized using the near-linear log-PCA method of \citet{Davies18} and \citet{Bosman22}, which reproduces the intrinsic continuum with $\sim 8\%$ accuracy and well-characterized wavelength-dependent uncertainties. All spectra are rebinned into $4\,h^{-1}\mathrm{Mpc}$ intervals to keep it consistent with the simulations, and non-detections are retained to preserve the noise statistics.

We apply the same masking strategy as in \citet{Spina25}, excluding rest-frame wavelengths $\lambda > 1185$\,\AA\ to avoid proximity-zone contamination, masking all damped Ly-$\alpha$\ systems (DLAs) identified in \citet{Davies24}, and removing all pixels within $\Delta v = 5000~\mathrm{km~s^{-1}}$ of the redshifted \ion{O}{VI}\,$\lambda\lambda1032,1038$\,\AA\ emission line to prevent spurious large-scale correlations. Residual sky-line contamination and all other cleaning steps described in \citet{Bosman22} are applied identically.

As discussed in \citet{Spina25}, we verified that the large-scale correlations observed in the data are not produced by continuum-reconstruction uncertainties. This was tested using 1000 mock realizations of the QSO sample, in which the continuum of each sightline is perturbed by coherent Gaussian offsets matched to the measured continuum uncertainties. These mocks reproduce the mild correlation seen at $z \lesssim 5.3$ but fail to generate the strong, extended correlations detected at higher redshift, even when the continuum-uncertainty amplitude is artificially increased by 50\%. This demonstrates that the correlations observed in the data reflect genuine large-scale structure in the IGM rather than systematics arising from continuum fitting.
\begin{figure*}
    \centering
    \includegraphics[width=0.96\textwidth]{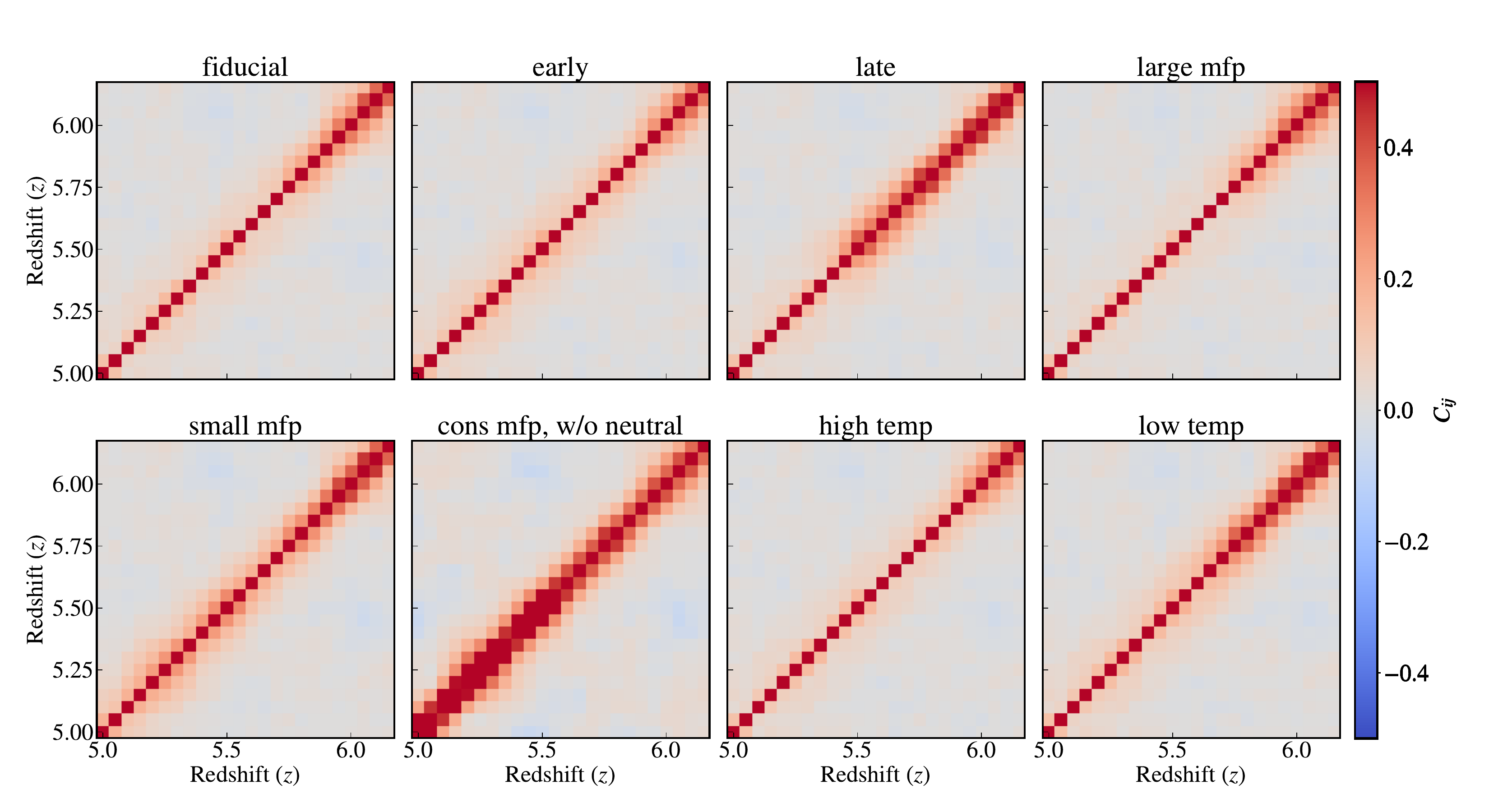}
    \caption{Correlation coefficients of the transmission flux between redshift ranges considered in this study ($z=5.0-6.1$). The panels show the different scenarios, as discussed in section \ref{sec:model_suite}. The correlation matrix derived from 67 quasar sigtlines (including E-XQR-30 samples) has been reported in \citet{Spina25}, which we show in Appendix \ref{app:appendix1}, for completeness.}
    \label{fig:corr_mat}
\end{figure*}
 \begin{figure*}
    \centering
    \includegraphics[width=\textwidth]{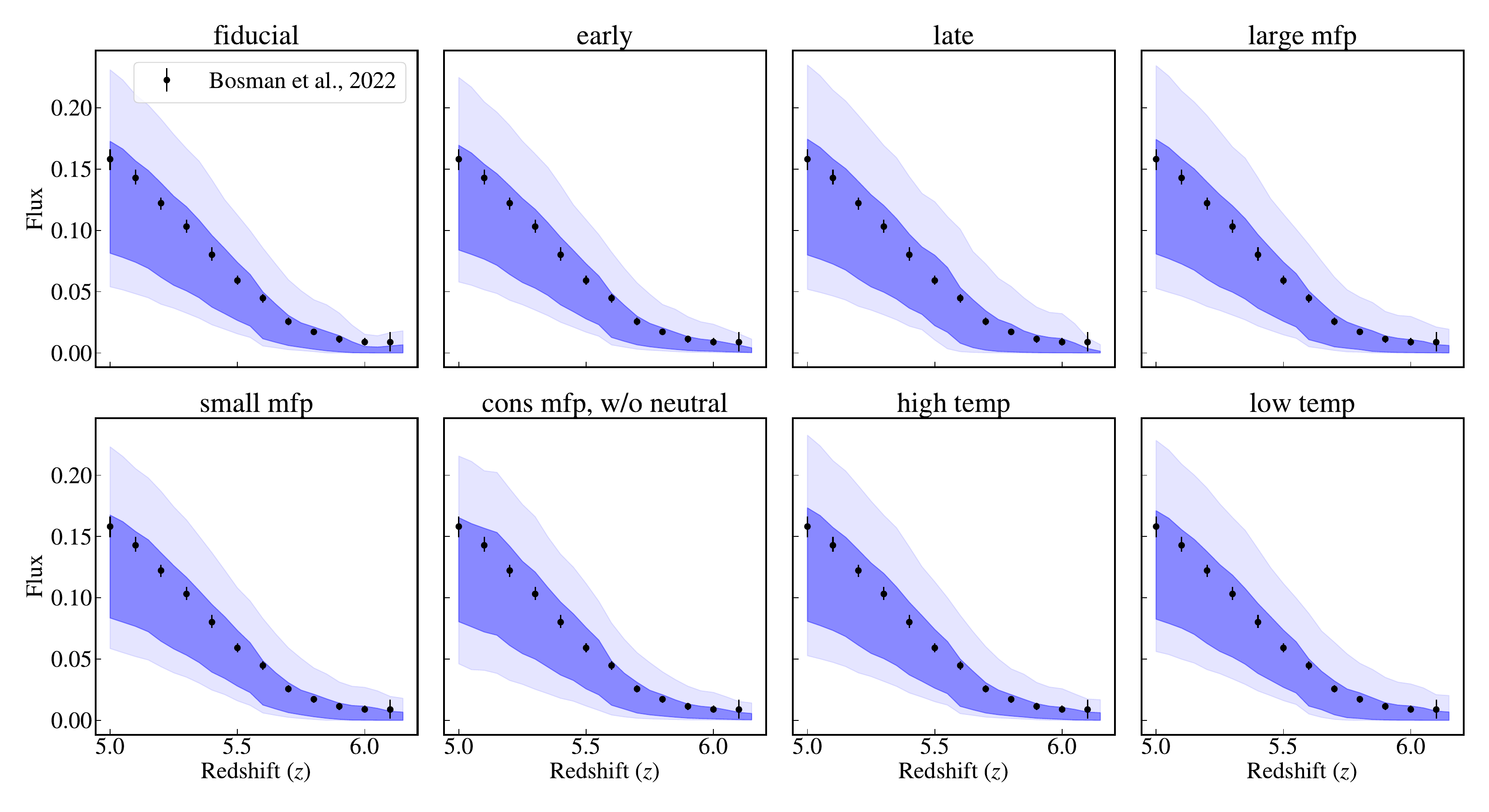}
    \caption{Flux distribution as a function of redshift, for the different scenarios as discussed in section \ref{sec:model_suite}. The bands show the 68\% and 95\% intervals of the flux distributions in the extracted skewers. The mean fluxes of the coeval simulation boxes have been matched with the observed mean, shown in black points \citep{Bosman22}.}
    \label{fig:flux_skew}
\end{figure*}

\section{Statistics with flux skewers}
\label{sec:stat}
We employ three complementary statistics in this work, each designed to capture large-scale information encoded in the Ly-$\alpha$ forest. The first two are the cumulative distribution function (CDF) of dark-gap lengths in different redshift intervals and the fraction of spectra (or skewers) containing long dark gaps as a function of redshift. These statistics are motivated by earlier observational analyses, e.g., \citet{Zhu_2021}. The third is the correlation-coefficient matrix of the Ly-$\alpha$ transmitted flux across redshift bins, recently introduced by \citep{Spina25}, which was also the main motivation for utilizing these large box simulations. We extract $N$ (=15000) random flux skewers from the lightcone volumes, covering a redshift range, $z=4.9-6.2$. These are divided into multiple realizations of 42 sets (to mimic the observed dataset), which are further used to compute the statistics. Specifically, we use 1000 realizations to compute the bootstrap uncertainties on the dark gap statistics. Furthermore, we remove all the dark skewers from the analysis (both data and models) that hit the boundary of the lightcone redshift ranges, to keep the estimators robust.  Below, we briefly introduce each of the statistics by summarizing the computation procedures.
\subsection{CDF of dark gap lengths} This statistic characterizes the average distribution of dark gaps within a given redshift interval. We define a pixel as dark if its transmitted flux falls below the threshold ($F\le0.05$ in this study). Consecutive dark pixels are grouped to form a dark gap, whose physical length is obtained by summing the number of pixels within the gap and weighing by the pixel resolution. Once we have the distribution of dark gaps, we assign them a central redshift value that corresponds to the middle of their extent. We then divide the distribution into four redshift bins with an interval of $\Delta z =0.2$, from $z=5.1$ to $z=5.9$. The uncertainties on these distributions are computed by bootstrapping the skewer realizations. We compare the CDFs of the dark gap lengths in these intervals utilizing our simulated spectra and the spectra from the  E-XQR-30 dataset. While computing the dark gaps from observational data, we follow the same procedure as we did for the simulation, which makes the comparison consistent with each other. Specifically, the observational spectra are averaged over $4~ h^{-1}\mathrm{cMpc}$ scale to match the resolution of our simulation. In the observed dataset, the non-detected pixels are assumed to be a part of dark gaps if they fall between two consecutive dark patches. Otherwise, we mask those pixels from the analysis procedures.
\subsection{Evolution of dark gap fraction} The redshift evolution of long dark gaps serves as a sensitive diagnostic of residual neutral patches and UVB-fluctuation strength at late times. To estimate the fraction of skewers exhibiting long dark gaps, we check for the presence of dark pixels which are part of a long gap ($L\ge 30 h^{-1}\mathrm{cMpc}$) in each redshift interval ($\Delta z$) of 0.02 and count the numbers. Then we divide the count of skewers providing long gaps by the total number of skewers in the corresponding range to estimate the fraction, $F_{30}$. We compute this fraction as a function of redshift, which provides a quantification of the dark gap frequencies with redshift evolution. Notably, we remove the possible contamination of damped Ly-$\alpha$ systems and metal absorbers from the dataset (as discussed in Section \ref{sec:obs_dat}) to make the comparison with simulations as consistent as possible.
\subsection{Correlation Matrix} This estimator reveals the large-scale correlation in the observed transmitted flux, which sets the motivation of the work. Similar to \citet{Spina25}, we bin the skewers with a redshift interval of $\Delta z=0.05$. If $F_{i,j}$ is the flux value for $i$-th binned skewer at redshift bin $j$, the covariance matrix is estimated as 
\begin{equation}
    \mathcal{S}_{ij} = \sum_{k=1}^{N}(F_{k,i}-\langle F\rangle_{i})^T(F_{k,j}-\langle F\rangle_{j})
\end{equation}
where $\langle...\rangle$ represents the average over all the skewers (i.e. $N=15000$ in our case). This further provides the correlation matrix as
\begin{equation}
    \mathcal{C}_{ij} = \frac{\mathcal{S}_{ij}}{\sqrt{\mathcal{S}_{ii}\mathcal{S}_{jj}}}
\end{equation}
Specifically, we look for the behavior of off-diagonal terms, where a strong positive enhancement would suggest correlated structures among redshift bins.\\
\subsection{Flux distribution} Lastly, we also utilize the skewers to estimate the transmitted flux distribution as a function of redshift. We compute the fluxes by averaging over a scale corresponding to $\Delta z=0.1$, keeping it consistent with the measurements by \citet{Bosman22}. This provides a qualitative estimate of the distribution width of fluxes at different redshifts.

\section{Results and interpretations}\label{sec:results}
In this section, we present our results and provide a qualitative discussion of how the different model variants impact the three statistics introduced earlier. 

In Figure \ref{fig:dark_cdf}, we show the CDFs of the dark gap lengths across different redshift intervals. The solid curves represent the distributions derived from the observed E-XQR-30 sightlines, while the dashed ones correspond to mean predictions from our model skewers. The shaded region shows the 68\% uncertainty in the distribution by sampling different model realizations. Overall, the fiducial model reproduces the observed trends well across all redshift bins. Specifically, at the highest redshift interval, $5.7\le z\le 5.9$, the model correctly mimics the distribution width towards large dark gap lengths ($L\ge 25 h^{-1}\mathrm{cMpc}$), although it slightly underpredicts the abundance of shorter gaps and overpredicts the longer gaps, corresponding to a normalized $L_1$ distance of $\sim0.053$ (see Appendix \ref{app:appendix3}\footnote{ we summarize $L_1$ metric comparison between models and data in Table \ref{tab:L1_metric}} for details) the data CDF. The relative lack of short length gaps at the corresponding redshift range points towards the lack of small neutral islands in the fiducial model. This is also evident from the "late" end model, where the relative number of short length gaps is larger, yielding an improved match with the data (normalized $L_1$ distance of $\sim0.030$). Not surprisingly, the deviation is slightly more pronounced for the "early" scenario due to a lower neutral fraction (normalized $L_1$ distance of $\sim0.065$). Similarly, the model with a lower temperature slightly improves the distribution at $5.7\le z\le 5.9$ by elevating the number of small dark gaps (normalized $L_1$ distance of $\sim0.046$). Not surprisingly, the higher temperature model degrades the match as it provides a slightly earlier reionization end and fewer dark gaps (normalized $L_1$ distance of $\sim0.067$).  The models with a larger or a smaller mean free path do not affect these statistics much, providing trends between the "fiducial" and "early" ones. However, a redshift evolving $\lambda_0$ is necessary to match the shape of the distribution at lower redshift intervals (i.e., $5.1\le z\le 5.7$). This can be seen in the "cons mfp, w/o neutral" case, where the model predicts too many dark gaps at lower redshifts, disfavoring the data distribution at the interval $5.1\le z\le 5.3$. The overall (combined for four redshift intervals, covering $5.1\le z\le 5.9$) normalized $L_1$ value for this model is the highest ($0.095)$ among the variants, while the "late" scenario provides the lowest ($0.065$).

Next, in Figure \ref{fig:F30_z}, we show the data and model comparison using $F_{30}$ statistics as defined earlier. The red lines correspond to the data distribution, while the dashed blue lines are the estimates from the models. The estimates of \citet{Zhu_2021} differ slightly (black dashed) from this work due to differences in the assumed pixel resolution of the flux skewers and the number of sources. For all the cases, the dark fraction value increases as we move towards higher redshifts, signifying the increased presence of large opaque regions. The data reveal the emergence of long dark gaps beginning at $z=5.3$, which is consistent with the recent findings of large-scale opacity fluctuations at those redshifts. We also find that the model variants of different reionization ends (i.e., "fiducial", "early", and "late") do not affect the distribution significantly and provide consistent trends with the derived estimates from observed quasar spectra. Likewise, models incorporating temperature changes or modifications to the photon mean free path do not produce significant deviations from the fiducial behavior, offering limited constraining power in these parameter spaces. We note that the data indicate a slight tension with respect to the mean predictions of the models at the redshifts $z\approx 5.3-5.5$, by producing an elevated long dark gap fraction. This excess may hint at the persistence of sizable neutral patches at these redshifts, although the current sample size limits the statistical significance of this feature. In agreement with the discussion in the previous paragraph, "cons mfp, w/o neutral" predicts a larger fraction of long dark gaps even at lower redshift ($z<5.3$) with a large uncertainty. This comes as a consequence of the assumed small (and constant) mean free path throughout the redshifts. However, as discussed earlier,  this scenario is already disfavored by the CDF distributions at lower redshifts. Overall, these estimators can not fully alleviate the degeneracy between the different models, providing a consistent picture with earlier studies \citep{Zhu_2021}.

We present the correlation matrix between the redshift bins in Figure \ref{fig:corr_mat}. This exploration is motivated by the observed large-scale correlation \footnote{\url{https://drive.google.com/file/d/1cR4YBad9tdOOxKaPp3vSfqh1cLPJhLUY/view?usp=share_link}} reported by \citet{Spina25}. Their measurements indicate that the characteristic correlation scale increases toward higher redshifts, reaching values as large as $\ge 150~h^{-1}\mathrm{cMpc}$. This corresponds to a redshift bin interval $\Delta z_{\mathrm{bin}}\gtrsim 0.3$ at a typical redshift of $z=5.7$. However, neither of our model variants is able to produce such a large correlation length. Although the fiducial scenario does exhibit a modest redshift evolution of the correlation, its amplitudes and scales fall well short of the observed estimates \citep{Spina25}. The correlation length scale increases a bit in the presence of the neutral islands for the late ending case, but that again is not sufficient to explain the observed correlation. Similarly, the usage of a short mean free path model slightly inflates the correlations across redshifts, but fails to match the data. These discrepancies suggest that a more comprehensive exploration of the model parameter space is required, possibly involving more sophisticated physical treatments. Fully numerical hydrodynamical simulations are the ideal avenue for such an investigation, although the box size requirement ($\gtrsim 500~h^{-1}\mathrm{cMpc}$) remains computationally prohibitive. However, we can still play with other inherent assumptions within the semi-numerical setup. For instance, our models assume a fixed fiducial power law dependence of the mean free path on density ($\propto \Delta^{-1}$) and photoionization rate ($\Gamma_{\mathrm{HI}}^{2/3}$), while generating the UVB fluctuations. We need to check how the variations on these relationships affect the large-scale correlations. Furthermore, the models approximate the radiative feedback effects by assuming a global minimum threshold mass ($M_{\mathrm{min}}$) for a step-like suppression, whereas in reality, this quantity can be spatially varying and the effect can be gradual. This may affect the large-scale morphology during reionization, although it is unlikely to produce an extremely large-scale correlation such as the observed ones. It is also worth noting that our models are based on a semi-numerical approach, where the small-scale astrophysical information is approximated via empirical scaling relations. The framework can be improved by synergising with the outputs of galaxy formation/evolution simulations. The other implications include cosmological modifications of the underlying density field, which may demand an involvement of non-standard physics at those redshifts. This may include primordial non-Gaussianities, modifications in the inflationary models, warm dark matter scenarios, or even modified general relativity. However, we caution the readers that one must explore more on the astrophysical front before relying on a cosmological solution. From the observational side, it is necessary to increase the number of quasar spectra for more robust estimates of the statistics and check the persistence of large-scale correlations utilizing larger samples. Specifically, we highlight that more samples will be crucial not only to get statistical distinctions among the models but also to strengthen the observed signature. For example, if the line of sight passes through proximity zones of any foreground sources, that may give rise to correlated fluxes. Ideally, it would also be useful to have an independent set of samples from a different observational facility that will give us confidence against any hidden observational systematics.

Finally, in Figure \ref{fig:flux_skew}, we show the flux distribution of the skewers as a function of redshift. It is apparent that all the models obey the observed mean transmission fluxes, which have been imposed by the construction of these model variants. The widths of the distributions are very similar, which makes it not suitable for distinguishing between models.

\section{Summary and conclusions}\label{sec:summary}
The final stages of the epoch of reionization remain an open question in modern cosmology. Recent observational evidence, particularly the detection of large-scale fluctuations in Ly-$\alpha$ forest opacity, appears to favor a relatively late end to reionization. However, sufficient room exists for a relatively early end, but with opacity fluctuations driven by spatial variations in the post-ionization UVB background. Motivated by the large-scale Ly-$\alpha$ transmission correlations reported by \citet{Spina25}, we employ a gigaparsec-scale simulation framework for modeling Ly-$\alpha$ forest opacity distributions, based on the efficient semi-numerical method of \citet{Maity_davies_2025}. Using this setup, we investigate the statistical properties of dark pixels in the Ly-$\alpha$ forest flux by comparing the observed high-redshift quasar spectra from the E-XQR-30 sample and the variants of our large simulation boxes. Our main conclusions are summarized below:
\begin{itemize}
    \item We constructed eight model variants (including the fiducial one) by varying different model ingredients such as mean free path, reionization end, and temperatures. The fiducial model was chosen such that it obeys a variety of recent observational constraints during the late phase of the EoR. In each case, the models were tuned to match the mean transmission fluxes at the redshift ranges ($5.0\le z\le6.1$) covered in this study. 
    \item We utilized these models to compare the CDF of the dark gap (defined as contiguous pixels with flux below 0.05) distributions with the samples from E-XQR-30 dataset. We found that our fiducial model provides a reasonable match to the data, while the variants with slightly late reionization end and lower temperatures improve the agreement. On the contrary, the data seem to disfavor the model with a short and constant mean free path without any presence of neutral island towards lower redshift bins ($z<5.5$).
    \item We further compared the fraction of skewers with long dark gaps ($\ge 30~h^{-1}\mathrm{cMpc}$) as a function of redshift. The model variants again provide excellent matches to the data. We found a slight tension in the redshift range $5.3-5.5$, where the data indicate a relatively large fraction of long dark gaps. However, this remains statistically inconclusive, and we need more samples for any further conclusions.
    \item Lastly, following our original motivation, we checked the correlation coefficients between the redshift bins. We found that none of our model variants produce an extremely large-scale correlation as reported by \citet{Spina25}. This suggests that the observed correlation may have cosmological implications beyond the standard realm. However, a more detailed exploration of the model parameters by alleviating astrophysical assumptions is necessary for any further conclusions. 
\end{itemize}
The study provides a robustness check of our large-scale Ly-$\alpha$ forest simulation model against the cutting-edge observations of dark gap distribution during late phase of reionization. In the future, the models will be useful to strategise large-scale surveys of high-redshift IGM.  
   
\section*{Data Availability}

The data presented in this article will be shared on reasonable request to the corresponding author (BM).

-------------------------------------------------------------------
\bibliographystyle{aa}
\bibliography{lya_opacity}

\begin{appendix}
\section{Observed correlation matrix}
\label{app:appendix1}
\begin{figure}
    \centering
    \includegraphics[width=\linewidth]{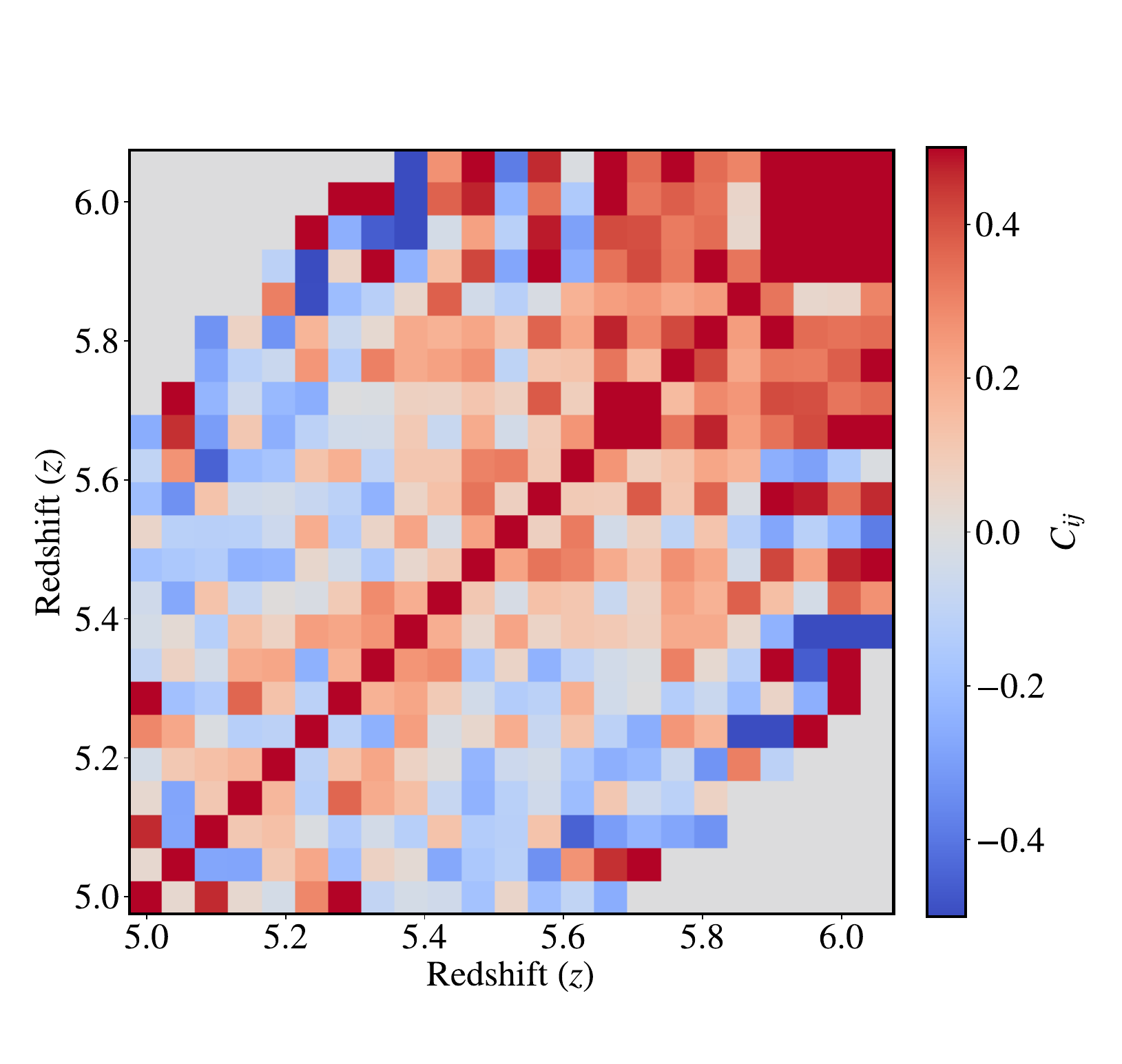}
    \caption{Observed Ly-$\alpha$ transmission flux correlation matrix in redshift range $5.0<z<6.1$, reproduced from \citet{Spina25}.}
    \label{fig:obs_corr}
\end{figure}

In Figure \ref{fig:obs_corr}, we show the correlation matrix derived from observed Ly-$\alpha$ transmission in the redshift range of $5.0<z<6.1$, as described in \citet{Spina25}. The large correlation values for the off-diagonal terms suggest that fluxes are strongly correlated across redshift bins. The previous study also quantified the scale of correlations, making it utilizable for model comparison. The study suggested a gradual increase in correlation length with increasing redshift, producing a value of $\sim150~h^{-1}\mathrm{cMpc}$ around $z\sim5.7$.

\section{Snapshots for the rest of the scenarios}\label{app:appendix2}
\begin{figure*}[t]
    \centering
    \includegraphics[width=\textwidth]{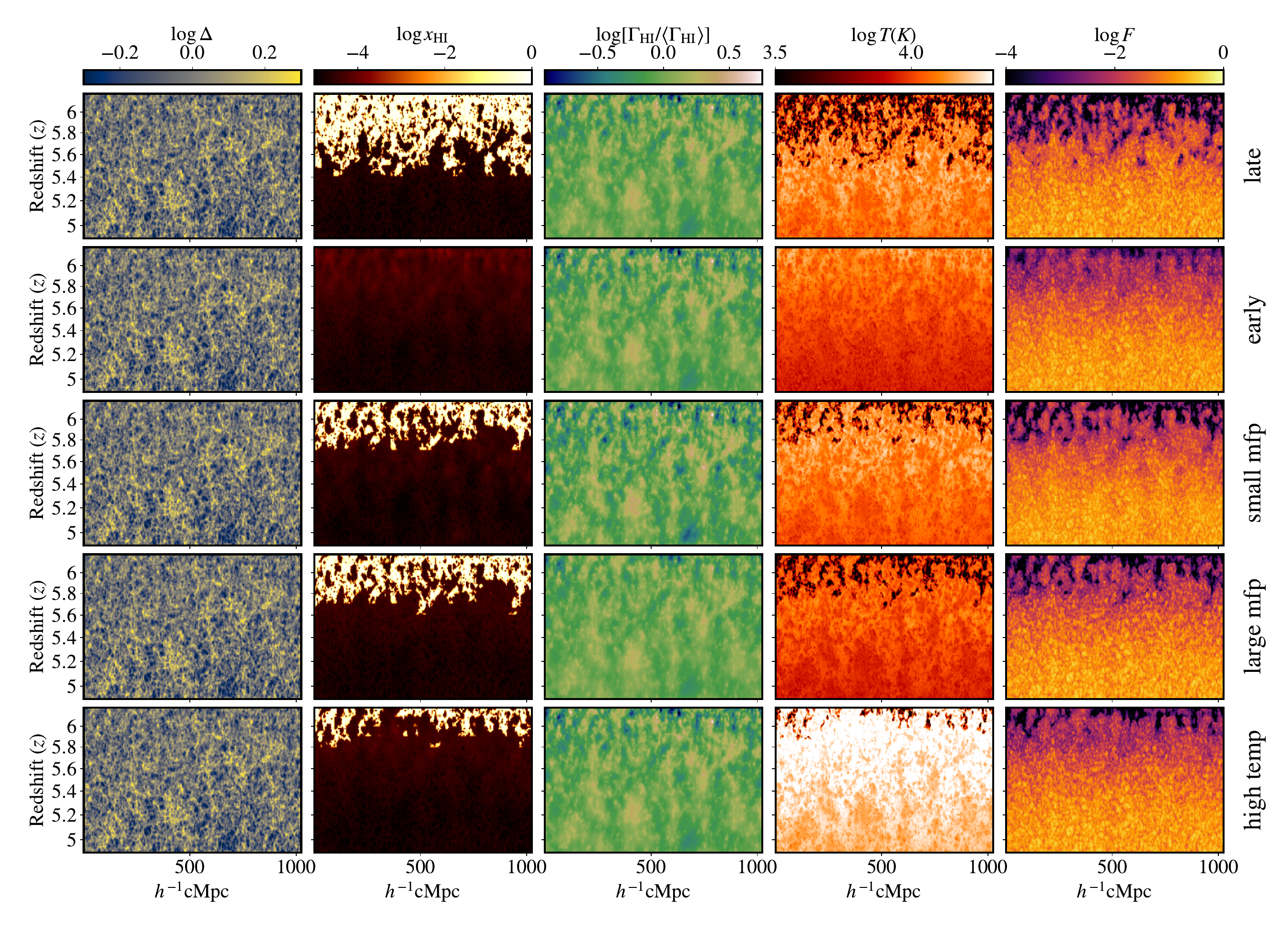}
    \caption{Lightcone snapshots for the rest of the five cases complementing Figure \ref{fig:mdel_snap} (i.e., "late", "early",  "small mfp", "large mfp" and "high temp", respectively from top to bottom). The columns correspond to density ($\Delta$), neutral fractions ($x_{\mathrm{HI}}$), UVB fluctuations ($\Gamma_{\mathrm{HI}}/\langle \Gamma_{\mathrm{HI}}\rangle$), temperature ($T$), and flux ($F$). The colorbars have been shown in logarithmic scales.}
    \label{fig:mdel_snap_appndix}
 \end{figure*}   
 
In Figure \ref{fig:mdel_snap_appndix}, we show the lightcone snapshots of the different physical quantities for the rest of the scenarios, complementing Figure \ref{fig:mdel_snap}. It is apparent that the "late" reionization end shows an abundance of neutral patches at a redshift as low as $z\sim5.4$, while the "early" scenario ends before $z\sim6.1$. Similarly, "small mfp" and "large mfp" provide slightly enhanced and reduced UVB fluctuations, respectively. The enhancement in temperature is also visually well distinguishable for the "high temp" model from the others. We further note that the mean flux evolution of all the lightcones has been matched with observational expectations by tuning the magnitude of the photoionization rate.

\section{Quantitative comparison of CDFs}\label{app:appendix3}
To provide a quantitative comparison among the CDFs of dark gap distribution, we estimate the $L_1$  distance between the data ($P_{\mathcal{D}}(<L)$) and model mean ($P_{\mathcal{M}}(<L)$). The metric is defined as 
\be
L_1 = \Delta x\sum_{i=1}^N\mid P_{\mathcal{D}}(<L) - P_{\mathcal{M}}(<L) \mid
\ee
, where $N$ is the number of bins ($=36$) and $\Delta x$ is the resolution ($=4~h^{-1}\mathrm{cMpc}$). In \tab{tab:L1_metric}, we show the normalized $L_1$ distance ($L_1$ divided by the total range to make it a dimensionless metric) between the CDF of the data and the mean CDF of various model suites. It is apparent that a lower value of this metric suggests a better resemblance between the model and the data distribution.
\begin{table*}
\centering
\caption{Normalized $L_1$ distance comparison between dark gap CDF data and the corresponding mean estimates for different models}\label{tab:L1_metric}
\renewcommand{\arraystretch}{1.3}
\setlength{\tabcolsep}{5.5pt}
\small
\begin{threeparttable}
\begin{tabular}{cccccc}
\hline

\textbf{Models}  & Normalized $L_1$  & Normalized $L_1$  & Normalized $L_1$  & Normalized $L_1$  & Normalized $L_1$  \\
 & ($5.1<z<5.3$) & ($5.3<z<5.5$) & ($5.5<z<5.7$) & ($5.7<z<5.9$) & \textbf{Total} \\
\hline

fiducial & $0.60/144$ & $1.20/144$ & $1.49/144$ & $7.67/144$& $10.96/144$\\
early & $0.85/144$ & $1.26/144$ & $1.53/144$ & $9.32/144$ & $12.97/144$\\
late & $0.48/144$& $1.19/144$ & $3.32/144$ & $4.40/144$ & $9.39/144$\\
large mfp & $0.41/144$ & $1.22/144$ & $1.48/144$ & $9.05/144$ & $12.17/144$\\
small mfp & $0.83/144$ & $1.23/144$ & $1.88/144$ & $8.38/144$ & $12.32/144$\\
cons mfp, w/o neutral & $2.16/144$ & $1.45/144$  & $3.40/144$ & $6.68/144$ & $13.69/144$ \\
high temp & $0.54/144$ & $1.22/144$ & $1.53/144$ & $9.71/144$ & $13.00/144$\\
low temp & $0.74/144$ & $1.20/144$ & $1.59/144$ & $6.64/144$ & $10.17/144$\\
\hline \\

\end{tabular}
\end{threeparttable}
\end{table*}
\end{appendix}
\end{document}